\documentclass[10pt, conference]{IEEEtran}
\IEEEoverridecommandlockouts
\usepackage{multirow,tabularx} 
\usepackage{adjustbox} 
\usepackage{booktabs}
\usepackage{color, colortbl} 
\usepackage{array, booktabs, makecell}
\usepackage{comment} 
\usepackage{color,soul}
\usepackage{graphicx}
\usepackage{caption}
\usepackage{subcaption}
\usepackage{nth}
\usepackage{url}
\usepackage{textcomp}
\usepackage[normalem]{ulem}
\usepackage{pgfplots}
\usepackage{tikz}
\usetikzlibrary{patterns}
\usepackage[figuresright]{rotating}
\usepackage[most]{tcolorbox}   
\usepackage{textcomp}  
\usepackage{mathtools}
\usepackage{tabularx}
\usepackage{pgfplots}
\usepackage{pgfplotstable}
\usepackage{sfmath}
\usepackage{array, booktabs, makecell}
\usepackage{color}
\usepackage{csquotes}
\usepackage{url}
\usepackage{comment}
\usepackage{tikz}
\usepackage{listings}
\usepackage{breqn}
\usepackage{booktabs} 
\usepackage{multirow}
\usetikzlibrary{fit} 
\usetikzlibrary{positioning}
\usetikzlibrary{arrows}
\usetikzlibrary{shapes.multipart}
\usepackage{float}
\usepackage{enumitem}
\usepackage[small,compact]{titlesec}
\usepackage{mathtools}
\usepackage{pgfplots}
\usepgfplotslibrary{fillbetween}
\pgfplotsset{compat=1.11,width=10cm}
\usepgfplotslibrary{statistics}

\usepackage{pgfplots, pgfplotstable}

\usepackage{bchart}
\usepackage[flushleft]{threeparttable}

\usepackage{pgf-pie}  
\usepackage{tabularx}
\usepackage{pgfplots}
\usepackage[nomessages]{fp}
\newcommand{\modulo}[2]{{#1}}
\usepackage{fontawesome}
\usepackage{calculator}
\usepackage{amsmath}

\usetikzlibrary{decorations.pathmorphing}
\tikzset{snake it/.style={decorate, decoration=snake}}
\usepackage{xcolor}
\definecolor{MyOrange}{RGB}{237,125,49}
\definecolor{MyBlue}{RGB}{91,155,213}
\usepackage{listings}

\definecolor{javared}{rgb}{0.6,0,0} 
\definecolor{javagreen}{rgb}{0.25,0.5,0.35} 
\definecolor{javapurple}{rgb}{0.5,0,0.35} 
\definecolor{javadocblue}{rgb}{0.25,0.35,0.75} 
 
\usepackage{draftwatermark}
\SetWatermarkText{PREPRINT}
\SetWatermarkLightness{0.88}
\SetWatermarkScale{0.33}

\definecolor{light-red}{rgb}{1,0.92,0.91}
\definecolor{light-green}{rgb}{0.9,1,0.93}
\definecolor{light-gray}{rgb}{0.95,0.95,0.95}
\newtcolorbox{box1}[1][]
{
  enhanced,
  sharpish corners,
  colback=blue!20,
  colframe=black!75,
  boxrule=1pt,
  top=0.05cm,
  bottom=0.05cm,
  left=0.05cm,
  right=0.05cm,
  enlarge top by=0.2cm,
  enlarge bottom by=0.2cm,
  breakable
}
\newtcolorbox{box2}[1][]
{
  title=#1, 
  enhanced,
  colback=blue!20,
  frame hidden,
  colbacktitle=gray, 
  boxed title style={colframe=gray},
  top=0.05cm,
  bottom=0.05cm,
  left=0.15cm,
  right=0.15cm,
  enlarge top by=0.5cm,
  enlarge bottom by=0.3cm,
  attach boxed title to top left={xshift=3mm,
  yshift=-3mm,yshifttext=-1mm},
  borderline west={4pt}{0pt}{blue!50!black},
  breakable
}
\usepackage[flushleft]{threeparttable}

\usepackage[frozencache,cachedir=.]{minted}
\setminted{bgcolor=light-gray}
\makeatletter
\patchcmd{\minted@colorbg}{\noindent}{\medskip\noindent}{}{}
\apptocmd{\endminted@colorbg}{\par\medskip}{}{}
\makeatother

\usepackage{tikz}

\def\BibTeX{{\rm B\kern-.05em{\sc i\kern-.025em b}\kern-.08em
    T\kern-.1667em\lower.7ex\hbox{E}\kern-.125emX}}
\begin{document}

\title{\huge
    Do the Test Smells \textit{Assertion Roulette} and \textit{Eager Test} Impact Students' Troubleshooting and Debugging Capabilities?
}

\author{\IEEEauthorblockN{Wajdi Aljedaani\IEEEauthorrefmark{1},
Mohamed Wiem Mkaouer\IEEEauthorrefmark{2}, Anthony Peruma\IEEEauthorrefmark{3} and
Stephanie Ludi\IEEEauthorrefmark{1}
}
\IEEEauthorblockA{\IEEEauthorrefmark{1}University of North Texas. Email\{wajdi.aljedaani, Stephanie.Ludi@unt.edu\}}
\IEEEauthorblockA{\IEEEauthorrefmark{2}Rochester Institute of Technology. Email\{mwmvse@rit.edu\}}
\IEEEauthorblockA{\IEEEauthorrefmark{3}University of Hawaii at Manoa. 
Email\{peruma@hawaii.edu\}}
}


 

\maketitle

\begin{abstract}
To ensure the quality of a software system, developers perform an activity known as unit testing, where they write code (known as test cases) that verifies the individual software units that make up the system. Like production code, test cases are subject to bad programming practices, known as test smells, that hurt maintenance activities. An essential part of most maintenance activities is program comprehension which involves developers reading the code to understand its behavior to fix issues or update features. In this study, we conduct a controlled experiment with 96 undergraduate computer science students to investigate the impact of two common types of test smells, namely \textit{Assertion Roulette} and \textit{Eager Test}, on a student's ability to debug and troubleshoot test case failures. Our findings show that students take longer to correct errors in production code when smells are present in their associated test cases, especially \textit{Assertion Roulette}. We envision our findings supporting academia in better equipping students with the knowledge and resources in writing and maintaining high-quality test cases. Our experimental materials are available online\footnote{https://wajdialjedaani.github.io/testsmellstd/}


\end{abstract}


\section{Introduction}
\label{sec:Introduction}

An essential activity in ensuring the quality of a software system is unit testing, where developers write code to verify the behavior of the implemented system's production (i.e., source) code \cite{pressman2019software}. By using  unit tests, organizations and project teams automate the discovery of flaws in their system that would otherwise go unnoticed or consume developer time \cite{hamill2004unit}. Given this invaluable benefit in improving the overall quality of a software system, many projects and organizations mandate that developers write unit tests as part of their software development process \cite{khorikov2020unit}, including the adoption of a test-driven development approach \cite{beck2003test}.

However, as with production code, test code is also subject to bad programming practices by developers, known as test smells \cite{van2001refactoring}. Likewise, similar to code smells, test smells are also an indicator of deeper problems, such as bad design or implementation choices in the test suite. Prior research has shown that test smells negatively impact the system's maintainability. Specifically, test smells have been shown to increase the change- and defect-proneness of the system's codebase \cite{Spadini2018ICSME}, increase the flakiness of test cases \cite{CamaraSAST21}, and negatively impact test code readability and understandability \cite{van2001refactoring}. Furthermore, developers' injection of these test smells ranges from lack of proper testing discipline (i.e.,  mistakes/carelessness and non-removal of debugging code) to testing knowledge gaps \cite{peruma2019CASCON}. 

As described above, through a series of empirical studies and developer interviews, the research community has shown that test smells impact a system's internal quality, thereby impacting maintenance activities. To further validate these findings and expand the body of knowledge on test smells, our study investigates the effect of test smells on code comprehension activities. To this extent, we conducted a sizeable human-based study with undergraduate students enrolled in a computer science program at a university in North America. 

Similar to code smells, there are multiple types of test smells defined in published literature \cite{aljedaani2021test}. However, as this is a human-based study and involves students, examining each smell in the test smells catalog is not feasible due to time constraints. Therefore, in this study, we focus our analysis on only two smell types-- \textit{Assertion Roulette} and \textit{Eager Test}. We arrived at these two smell types by reviewing multiple studies \cite{bavota2012empirical,grano2019scented,peruma2019CASCON,Spadini2020MSR} and comparing the distribution of smell types that these studies have in common; both of these smell types frequently occur in the test suites of open-source Java systems. 

\subsection{Motivation \& Goal}

While there exist studies that evaluated the impact of test smells on the code comprehension capabilities of students, these studies either involved students writing complete test cases \cite{bai2022check,bai2021students,buffardi2021unit} or evaluating the test suites from large, well-established open-source systems with which they have no prior experience \cite{bavota2015test}. In contrast, as we elaborate in Section \ref{sec:ResearchMethod}, our work involves students examining pre-written test cases corresponding to simplistic use cases and making updates to the production (i.e., system under test) code to correct failing test cases. Hence, to a large extent, we eliminate any influence placed on the student's cognitive load caused by understanding (or being overwhelmed by) the overall behavior and technical design/architecture of a complex and unfamiliar system. Therefore, the findings from our study are more closely aligned with the actual impact of test smells on code comprehension. Furthermore, this study also allows us to compare our findings against the work by Bai et al. \cite{bai2022assertion}, which states that \textit{Assertion Roulette} should not be considered a bad smell for students.

In this study, our goal is to determine \textit{the extent to which the presence of test smells in the test suite impacts a student's troubleshooting and debugging capabilities}. Specifically, our work compares the effect that smelly and non-smelly test suites have on students when tasked with fixing failing test cases by only correcting defects in the production code of a system they are familiar with. We theorize that test files exhibiting test smells cause an increase in code comprehension time than those without test smells, resulting in students spending more time on maintenance activities.

\subsection{Contribution}
The results of our study show that test smells negatively impact a student's code comprehension capabilities. Furthermore, the \textit{Assertion Roulette} smell causes students to take more time to address issues in the production code than the \textit{Eager Test} smell. Our study highlights the need for academia to invest in and prioritize teaching students about all types of test smells, their harmful impact on maintenance activities, and tools that can be used to automatically detect and eliminate these smells from the test suite of a software system.

\section{Test Smell Definitions \& Related Work} 
\label{sec:RelatedWork}
This section provides definitions of the two test smell types (i.e., \textit{Assertion Roulette} and \textit{Eager Test}) utilized in our study and an overview of the related work in this area.

\subsection{Test Smell Definitions}

\noindent{\textit{Assertion Roulette.}} The passing/failing of a test case is determined by the execution of the assertion method it contains. These assertion methods allow developers to include an optional textual message indicating the reason for the failure of the assertion. This smell occurs when a test method contains two or more assertions without an explanation message. Troubleshooting the failure of a test case becomes challenging as the developer is unaware of the cause of the failure. 

\noindent{\textit{Eager Test.}} This smell occurs when a test method verifies multiple functionalities of the production code by invoking several production methods. This smell makes it hard to understand the true purpose of the test. Furthermore, it increases the coupling between the test method and production code, which, in turn, negatively impacts maintenance.

\subsection{Related Work}

Several automated techniques and tools for detecting test smells have been published in the literature \cite{aljedaani2021test}. In addition, researchers have identified a collection of test smells \cite{garousi2018smells}, while others have concentrated on the effects of test smells and removal techniques \cite{kim2021secret}. For example, Van Bladel and Demeyer suggested eliminating test smells in the context of refactoring test code \cite{van2017test}, and Van Deursen et al. highlighted harmful test smells and techniques to eliminate them \cite{van2001refactoring}. In addition, they offered conceptual and technical explanations for evaluating students' activity by identifying test smells in their code and offering test smell-related observations. This section highlights several prior studies that particularly shaped our methodology. Next, we divide the related work into two different aspects of test smell in education: software testing in education, where we focus on current approaches used to examine the testing in education; Students' programming and testing activities, which focus particularly on unit testing inside the classroom. Finally, Table~\ref {tab:RelatedWork} presents a summary of the systematic analysis studies in the related work.

\begin{table*}[t]
\centering

\caption{Summary of the systematic analysis studies in related work.}
\begin{adjustbox}{width=1.0\textwidth,center}
\begin{threeparttable}
\begin{tabular} {|c|c|l|c|c|c|c|}\hline

\rowcolor{gray!60} 
{\textbf{Study}} & {\textbf{Year}}   
& {\textbf{Purpose}} & {\textbf{Evaluation}} & {\textbf{Test Smell}}& {\textbf{Participant}} & {\textbf{\# of Participants}}\\ \hline
         &     &     &    &  TCD, MG, GF, ET,   &  Students   & 49 \\ \cline{6-7}
 \cite{bavota2015test}        &  2015   &  Understanding how test smells is spread in real software systems   &  Experiment  &  LT, AR, IT, SE  &  Developers   & 12 \\\hline
 
\rowcolor{gray!25}
 \cite{fraser2019gamifying}        &  2019   & Engaging students with software testing in an entertaining way    &  Survey  &  Game experiment  &  Students   & 123 \\\hline
 \cite{buffardi2021unit}        &  2021   &  Exploring smells exhibited by students first learning how to unit test   &  Unit tests, source code  &  MA, CL &  Students   & 246 \\\hline
\rowcolor{gray!25}
 \cite{bai2021students}        &  2021   &  Discovering students’ perceptions \& challenges when practicing unit testing   &  Survey  &  N/A  &  Students   & 54 \\\hline
 \cite{bai2022check}        &  2022   & Assessing the impact of the testing checklist    &  Survey  &  N/A  &  Students   & 32 \\\hline
\end{tabular}

\begin{tablenotes}
            \item \textbf{Abbreviation of test smells types:} (AR) Assertion Roulette, (MG) Mystery Guest, (GF) General Fixture, (ET) Eager Test, (LT) Lazy Test, (IT) Indirect Testing, (SE) Sensitive Equality, (TCD) Test Code Duplication, (MA) Multiple Assertions, (CL) Conditional Logic.
        \end{tablenotes}
\end{threeparttable}

\end{adjustbox}
\label{tab:RelatedWork}
\end{table*}

\subsubsection{Software Testing in Education}

Educators have examined a variety of assessment integration strategies for computer science courses. For instance, some researchers instruct students to submit their software tests and solutions \cite{edwards2012running, goldwasser2002gimmick}, while others involve students in peer testing \cite{roberge1994using, gaspar2013preliminary}. Fraser et al. \cite{fraser2019gamifying} proposed a Code Defenders game that is utilized to engage students' activities in the test suite. Aniche and colleagues \cite{aniche2019pragmatic} conducted a survey involving 84 first-year computer science students regarding the difficulties associated with learning software testing. They also investigated the errors that were made in the lab work of 230 students. According to their findings, there are eight different types of typical errors. These include test coverage, maintainability of test code, understanding of testing principles, boundary testing, state-based testing, assertions, mock objects, and tools. 

Previous studies have investigated both the level of quality of student-written test code \cite{aniche2019pragmatic, carver2011evaluating, edwards2003improving, edwards2014student} and the viewpoints of students on unit testing \cite{aniche2019pragmatic, gaspar2013preliminary}. These studies utilize several metrics, such as the frequency of defects identified by student-created test cases and branch coverage. To evaluate the incremental testing procedures of software development projects, Kazerouni et al. \cite{kazerouni2019assessing} developed new metrics: the balance and sequencing of the testing effort. According to an evaluation conducted by Carver and Kraft \cite{carver2011evaluating}, students in the senior year of computer science do not have the skills necessary to make good use of test-coverage techniques. These experiments were performed in a classroom setting, and participants were given grades for their participation.


\subsubsection{Students' programming and testing activities}

Several approaches \cite{aaltonen2010mutation, edwards2004using} have been developed for assessing student-created test suites. Bai et al. \cite{bai2022check} studied the impact of a checklist on the students writing test cases. The students anticipated that they would develop JUnit tests to check the functionality of a program that had been implemented to evaluate the student completeness, effectiveness, and maintainability. In recent work, Buffardi and Aguirre-Ayal \cite{buffardi2021unit} analyzed students' work on testing assignments to examine their adoption of test smells. The authors also investigated the relationship between three types of test smells and the test accuracy of the students' work. Bai et al. \cite{bai2021students} performed an experimental study to learn how students understand unit testing and what obstacles they face when engaging in unit testing. Bavota et al. \cite{bavota2015test} performed a control experiment on students and industrial developers on six test smells. They execute software comprehension tasks on test suites with and without test smells and measure performance using correctness and time. Participants cannot conduct the necessary maintenance when this test smell is present. Thus, the smell significantly affects it (bachelor students scored 48\% correctness, while industrial developers scored 42\%). 

Researchers and educators commonly use test case/suite success rates to evaluate the quality of student-written source code \cite{bai2019exploring, kazerouni2017quantifying, williams2001support}. In contrast, students' productivity is generally measured in lines of code per hour \cite{baheti2002exploring} or work session \cite{kazerouni2017quantifying}. Unfortunately, there has been a lack of focus on the importance of test smells in the classroom. In a study similar to ours, Bai et al. \cite{bai2022assertion} examined how the \textit{Assertion Roulette} smell affects students' productivity and conduct while writing code. The authors employed the Bowling Score Keeper project, and the student was tasked with writing a Java app to calculate the score of a single bowling game according to a set of specifications and JUnit tests. 

\section{Study Design}
\label{sec:ResearchMethod}

The primary objective of this study is to assess to what extent \textit{Assertion Roulette} and \textit{Eager Test} hinder students' program comprehension. To do so, we evaluate the impact of these smells' existence on the debugging process when students use smelly test files to locate errors. Therefore, our main Research Question is:


\vspace{-0.5cm}
\begin{box2}
\textbf{RQ: To what extent do \textit{Assertion Roulette} and \textit{Eager Test} impact the time spent by students in debugging failing test cases?}
\end{box2}
\vspace{-0.4cm}

By \textit{debugging time}, we mean the time needed for a student to troubleshoot a failing test method and fix its corresponding error in the production method under test. Based on existing studies, smelly test files hinder program comprehension, i.e., we hypothesize that students should take a longer time to locate an error raised by a smelly test method, as it is harder to read, in comparison to the time needed to find an error raised by a non-smelly test method. To address this hypothesis, we create a controlled experiment where we select one project, which already contains a unit test suite with 100\% path coverage. Then, we inject errors in production methods, which are going to be caught by the tests, i.e., for each error injected into the production code, its corresponding test method is going to fail. For a set of created errors, the time needed to debug their failing test methods is expected to take longer if these test methods are smelly. This can be empirically validated if, for the same set of errors, the time needed to debug them would take longer if the test suite is smelly, in comparison with the debugging time when the same test suite is not smelly. Since we are interested in comparing the \textit{debugging time}, out of a non-smelly test suite (referred to as \textit{Suite N}), we create two variants of the same test suite: The first variant (referred to as Suite A) has the same testing logic of Suite N, but with test methods infected with the \textit{assertion roulette} smell. Similarly, the second variant (referred to as \textit{Suite E}) has \textit{Suite N}'s methods infected with the \textit{eager test}. These suites are described as follows:


\begin{itemize}
  \item \textbf{Suite N}. It encompasses \textit{(N)on-Smelly} test cases. Each production method is associated with one or multiple test methods, testing multiple scenarios and ensuring the coverage of all the method's execution paths. We followed the guidelines by XUnit \cite{meszaros2007xunit}.
  
 \item \textbf{Suite A}. We introduced \textit{(A)ssertion Roulette} smell into \textit{Suite N} test cases by testing multiple scenarios, for one given production method, under the same test method. This induces a test method with multiple assert statements, covering all the production method's execution paths. According to the literature, it hinders comprehension by 
  making it difficult to determine which \textit{assertion} has triggered the test failure \cite{panichella2022test}.

  \item \textbf{Suite E}. We introduced \textit{(E)ager Test} smell into \textit{Suite N} test cases by testing multiple scenarios, for multiple production methods, under the same test method. This induces a test method with multiple assert statements, covering all the production method's execution paths. According to the literature, it hinders comprehension by 
  making it difficult to determine which \textit{method} has triggered the test failure \cite{panichella2022test}.
 
\end{itemize}

To ensure each suite contains (or not) the intended test smell type, we run TS-Detect, one of the popular state-of-the-art test smell detectors \cite{peruma2020tsdetect}, for each suite, as a sanity check.



\subsection{Project Overview}


Since we are measuring students' debugging time, we need to avoid any bias that can be introduced due to misunderstanding the program's behavior. Therefore, the chosen application should be intuitive for anyone to understand. For this purpose, we created a basic calculator application using Java programming language. The language was chosen to match students' familiarity with object-oriented programming at their level. Similarly, the choice of calculator is driven by intuitiveness and students' familiarity with its features under test. The calculator was designed with eight functions listed below:


\begin{itemize}
  \item \textbf{Summation (Sum):} is a production method that takes as input two variables of type \textit{double}, and returns their summation, 
  e.g., $ 10 + 2 = 12 $.
  
  \item \textbf{Subtraction (Sub):} is a production method that takes as input two variables of type \textit{double}, and returns their subtraction, e.g., $10 - 2 = 8$.
  
  \item \textbf{Multiplication (Mult):} is a production method that takes as input two variables of type \textit{double}, and returns their multiplication, e.g., $ 10 \times 2 = 20 $. 
  
  \item \textbf{Division (Div):} is a production method that takes as input two variables of type \textit{double}, and returns their division, e.g., $ 10 \div 5 = 2$. 
  
  
  \item \textbf{Square Root (SQRT):} is a factor that, when multiplied by itself, equals the original value of the given integer. A square root is represented by a radical symbol ($\sqrt{~~}$) and can be determined by the value of the power $\frac{1}{2}$ of an integer, e.g., $\sqrt{25}=5$.  
  
  \item \textbf{Modulo (Mod):} is the signed residue of a division, which occurs after dividing two numbers. It is computed by subtracting the divisor from the dividend until the resulting is less than the divisor, e.g., $5 \pmod{2}=\modulo{1}{}$.
  
  \item \textbf{Average (Avg):} is a mathematical operation to compute the mean of a given set of numbers. It is the ratio of sum of all numbers in a given set to the number of values present in the set, e.g., avg of numbers present in set $A= \{1,2,3,4,5\}$ can be computed as 
  $\frac {1+2+3+4+5}{5}=3 $. 
  
  \item \textbf{Factorial (Fact):} is a function that outputs the product of all positive integers less than or equal to a given positive integer. It is indicated by an exclamation mark preceding that integer, e.g., 4! = 24. 
\end{itemize}

\begin{table}
\centering
\caption{
Summary of test cases for each category.}
\label{tab:TestCases}
\begin{adjustbox}{width=0.5\textwidth,center}
\begin{tabular}{|c|c|c|c|}\hline

\rowcolor{gray!60}
  \multicolumn{1}{|c}{\cellcolor{gray!60}} &
  \multicolumn{3}{|c|}{\cellcolor{gray!60}\textbf{\# of Test Cases}} \\ \cline{2-4}
\rowcolor{gray!60}
  \multicolumn{1}{|c}{\multirow{-2}{*}{\cellcolor{gray!60}\textbf{}}} &

  \multicolumn{1}{|c}{\cellcolor{gray!60}\textbf{Suite N}} &
  \multicolumn{1}{|c}{\cellcolor{gray!60}\textbf{Suite A}} &
  \multicolumn{1}{|c|}{\cellcolor{gray!60}\textbf{Suite E}}  \\
\rowcolor{gray!60}
  \multicolumn{1}{|c}{\multirow{-2}{*}{\cellcolor{gray!60}\textbf{Method}}} &

  \multicolumn{1}{|c}{\cellcolor{gray!60}\textbf{Non-Smelly test}} &
  \multicolumn{1}{|c}{\cellcolor{gray!60}\textbf{Assertion Roulette}} &
  \multicolumn{1}{|c|}{\cellcolor{gray!60}\textbf{Earger Test}}  \\ \hline
    
  \multicolumn{1}{|c}{\textbf{Summation (Sum)}} &
  \multicolumn{1}{|c}{5} &
  \multicolumn{1}{|c}{1} &
  \multicolumn{1}{|c|}{\cellcolor{gray!30}\textbf{ }}  \\ \cline{1-3}
  \multicolumn{1}{|c}{\cellcolor{gray!30}\textbf{Subtraction (Sub)}} &
  \multicolumn{1}{|c}{\cellcolor{gray!30}5} &
  \multicolumn{1}{|c}{\cellcolor{gray!30}1} &
  \multicolumn{1}{|c|}{\cellcolor{gray!30}}  \\ \cline{1-3}
  \multicolumn{1}{|c}{\textbf{Multiplication (Mult)}} &
  \multicolumn{1}{|c}{7} &
  \multicolumn{1}{|c}{1} &
  \multicolumn{1}{|c|}{\cellcolor{gray!30}\textbf{ }}  \\ \cline{1-3}
  \multicolumn{1}{|c}{\cellcolor{gray!30}\textbf{Division (Div)}} &
  \multicolumn{1}{|c}{\cellcolor{gray!30}5} &
  \multicolumn{1}{|c}{\cellcolor{gray!30}1} &
  \multicolumn{1}{|c|}{\cellcolor{gray!30}9 }  \\ \cline{1-3}
  \multicolumn{1}{|c}{\textbf{Square Root (SQRT)}} &
  \multicolumn{1}{|c}{5} &
  \multicolumn{1}{|c}{1} &
  \multicolumn{1}{|c|}{\cellcolor{gray!30}}  \\ \cline{1-3}
  \multicolumn{1}{|c}{\cellcolor{gray!30}\textbf{Modulo (Mod)}} &
  \multicolumn{1}{|c}{\cellcolor{gray!30}5} &
  \multicolumn{1}{|c}{\cellcolor{gray!30}1} &
  \multicolumn{1}{|c|}{\cellcolor{gray!30}}  \\ \cline{1-3}
  \multicolumn{1}{|c}{\textbf{Average (Avg)}} &
  \multicolumn{1}{|c}{4} &
  \multicolumn{1}{|c}{1} &
  \multicolumn{1}{|c|}{\cellcolor{gray!30}}  \\ \cline{1-3}
  \multicolumn{1}{|c}{\cellcolor{gray!30}\textbf{Factorial (Fact)}} &
  \multicolumn{1}{|c}{\cellcolor{gray!30}4} &
  \multicolumn{1}{|c}{\cellcolor{gray!30}1} &
  \multicolumn{1}{|c|}{\cellcolor{gray!30} }  \\ \hline
  \multicolumn{1}{|c}{\cellcolor{blue!30}\textbf{Total}} &
  \multicolumn{1}{|c}{\cellcolor{blue!15}\textbf{40}} &
  \multicolumn{1}{|c}{\cellcolor{blue!15}\textbf{8}} &
  \multicolumn{1}{|c|}{\cellcolor{blue!15}\textbf{9 }}  \\ \hline
\end{tabular}%
\end{adjustbox}
\end{table}

\subsection{Test Suites Creation}

For each production method, we need to create its corresponding test methods, ensuring 100\% path coverage. These test methods were automatically generated by EvoSuite\footnote{https://www.evosuite.org/} and labeled as \textit{Suite N}. Since the generated test methods' names are not descriptive, for each test method, we added a comment to indicate which production method it tests. It is critical for our experiments to ensure that the mapping between test and production methods is maintained. Otherwise, the overhead of students searching for such mappings would inflate the debugging time. To create \textit{Suite A} and \textit{Suite E}, we duplicate \textit{Suite N}, and we manually introduce the smells based on their definitions that we outlined above.

To illustrate how these suites differ, Listing \ref{Listing:Group_A} shows 4 test methods from \textit{Suite N}. Listing \ref{Listing:Group_B} shows how \texttt{test00()} and \texttt{test01()} (resp. \texttt{test05} and \texttt{test06}) are merged, since they are testing the same production summation (resp. subtraction) method. The merged methods have the \textit{assertion roulette} smell, so they belong to \textit{Suite A}. As for Listing \ref{Listing:Group_C}, all methods from Listing \ref{Listing:Group_A} are merged into one test method, constituting the \textit{eager test}. The merged method has the \textit{eager test} smell, so it belongs to \textit{Suite E}. This process has resulted in 40 test methods in \textit{Suite N}, 8 test methods in \textit{Suite A}, and 9 test methods in \textit{Suite E}. The count of test methods for each test suite is summarized in Table \ref{tab:TestCases}

\subsection{Errors Generation}

To create the errors, we used PITest\footnote{https://pitest.org/}, a Java mutation testing framework, to generate faults (or mutations) that are purposely seeded into the production methods. For a given mutated production method, if one or many of its corresponding test method(s) fail(s), then the error is caught. Otherwise, if the tests pass, then the error is missed. PIT is typically used to evaluate the quality of tests by the percentage of caught errors. In our context, we use PIT to generate arbitrary errors throughout the production methods. Then we selected errors while making sure each production method would have at least one error, and the selected errors were all caught by the 3 test suites.  


Figure \ref{Listing:MethodsExamples} presents two production methods, i.e., summation and subtraction, containing two errors. In the $sum\_$ method, the summation operation (+) is replaced with the subtraction operation (-), resulting in a faulty behavior. Likewise, the subtract operation (-) is replaced with the summation operation (+), in the $diff\_$. 



\begin{lstlisting}[caption=Example for two production methods with seeded errors.,label=Listing:MethodsExamples, firstnumber = last, escapeinside={(*@}{@*)},escapechar=!]
public class Calculator {
    public double sum(double [] arr){
        //Creation of Array
        double sum_ = 0;
        for(int i =0; i < arr.length; i++){
            sum_ -= arr[i];
        }
        //adding all elements in an array
        System.out.println("Addition: "+sum_);
        return sum_;
    }

    public double subtract(double [] arr){
        //Creation of Array
        double diff_ = 0;
        for (int i = 0; i < arr.length;  i++){
            diff_ += arr[i];
        }
        //Subtracting all elements in an array
        System.out.println("Subtraction: "+diff_);
        return  diff_;
    }
}    
\end{lstlisting}

\subsection{Target Course}

This experiment has been conducted in an undergraduate senior-level software engineering course\footnote{Some details revealing the identity of the course's institution has been omitted for double-blind review.}. 
Before joining this course, students have about two years of programming experience. This provides them with the background to perform the debugging needed in this experiment. Also, they are familiar with the process of searching for the root errors in a faulty code.




\begin{lstlisting}[caption=Example of test case source code for Non-test Smell test suite.,label=Listing:Group_A, firstnumber = last, escapeinside={(*@}{@*)},escapechar=!]
public class Calculator_ESTest extends Calculator_ESTest_scaffolding {
    //Test Cases for Sum Method
    @Test(timeout = 4000)
    public void test00() throws Throwable {
        Calculator calculator0 = new Calculator();
        double[] doubleArray0 = new double[2];
        double double0 = calculator0.sum(doubleArray0);
        assertEquals(0.0, double0, 0.01);
    }
    @Test(timeout = 4000)
    public void test01() throws Throwable {
        Calculator calculator0 = new Calculator();
        double[] doubleArray0 = new double[2];
        doubleArray0[0] = (-1.0);
        double double0 = calculator0.sum(doubleArray0);
        assertEquals((-1.0), double0, 0.01);
    }
    //Test Cases for Subtract Method
    @Test(timeout = 4000)
    public void test05() throws Throwable {
        Calculator calculator0 = new Calculator();
        double[] doubleArray0 = new double[2];
        double double0 = calculator0.subtract(doubleArray0);
        assertEquals(0.0, double0, 0.01);
    }
    @Test(timeout = 4000)
    public void test06() throws Throwable {
        Calculator calculator0 = new Calculator();
        double[] doubleArray0 = new double[5];
        doubleArray0[2] = 93.0;
        double double0 = calculator0.subtract(doubleArray0);
        assertEquals((-93.0), double0, 0.01);
    }
}   
\end{lstlisting}

\subsection{Pilot Study} 

A pilot study is the initial phase of the entire research protocol and is generally a smaller-scale study that serves to solidify the main study \cite{in2017introduction}. Therefore, prior to the primary investigation, we conducted a pilot study with four undergraduate students who were later excluded from the controlled experiment. The goal of the pilot study was to guarantee that the experiment's instructions were clear and to establish an approximate duration for the lab session. Following the pilot study, we decided to provide all upcoming participants with documentation on how to set up the programming environment. Also, we would only allow students to participate in the experiment when they have their environment ready to avoid skewing our measurements.

\begin{lstlisting}[caption=Example of test case source code for Assertion Roulette test suite.,label=Listing:Group_B, firstnumber = last, escapeinside={(*@}{@*)},escapechar=!]
public class Calculator_ESTest extends Calculator_ESTest_scaffolding {
    //Test Cases for Sum Method
    @Test(timeout = 4000)
    public void test00() throws Throwable {
        Calculator calculator0 = new Calculator();
        double[] doubleArray0 = new double[2];
        double double0 = calculator0.sum(doubleArray0);
        assertEquals(0.0, double0, 0.01);

        double[] doubleArray1 = new double[2];
        doubleArray1[0] = (-1.0);
        double double1 = calculator0.sum(doubleArray1);
        assertEquals((-1.0), double1, 0.01);
    }
    //Test Cases for Subtract Method
    @Test(timeout = 4000)
    public void test01() throws Throwable {
        Calculator calculator0 = new Calculator();
        double[] doubleArray0 = new double[2];
        double double0 = calculator0.subtract(doubleArray0);
        assertEquals(0.0, double0, 0.01);

        double[] doubleArray1 = new double[5];
        doubleArray1[2] = 93;
        double double1 = calculator0.subtract(doubleArray1);
        assertEquals((-93), double1, 0.01);
    }
}
\end{lstlisting}

\subsection{Procedure}

Following the results of the pilot study, we conducted two sessions: a preparation session and a controlled experiment session. During the preparation session, we supplied students with a video tutorial for both Windows and Mac to show them how to install setup the programming environment on their computers and run test cases on their IntelliJIDEA\footnote{https://www.jetbrains.com/idea/} panels. We also gave the students written instructions outlining a step-by-step procedure for installing Java on their systems\footnote{These instructions are included in our replication package}. We made sure all students had their environment ready and knew how to run test cases prior to our experiment. Also, we provided students with the bug-free version of the project, along with some test cases from \textit{Suite N}, as we want to increase student's familiarity with the production methods. Students' familiarity with the production code is important, as some students may exercise more effort to read and comprehend source code \cite{yenigalla2016novices}. Code comprehension is also a noise that we mitigate through their exposure to the project before the session. The supporting material related to the current experiment can be accessed through the link: \cite{ProjectWebiste}. Finally, we conducted a presentation of the calculator project, its features (operations), its source methods, and the execution of sampled test cases.


The second session was carried out in person to avoid any collusion. At the start of the controlled experiment session,  we randomly split students into three groups, based on which test they will be using: \textit{N}, \textit{E}, and \textit{A}. The entire session was 2 hours (120 minutes) long. The experiment started at the same time for every student. The task assigned to the students was to identify and fix the issues raised by the failing test methods in their corresponding suite. The use of online resources was also permitted. Students were instructed to submit their updated code immediately to Canvas\footnote{Web-based learning management system. https://instructure.com/canvas} once they are done fixing the errors. We chose to use Canvas since students are familiar with it. We determined \textit{debugging time} for each student by examining their submission timestamp on Canvas. Finally, we shared an online post-survey to gather their feedback about their debugging experiences. We use this survey to gauge if students have experienced any difficulties when debugging their code. It is important to note that students are not aware of the underlying experiment, i.e., the multiple test suites and the existence of test smells.



\begin{figure}[t]
   \centering
   \includegraphics[width=0.5\textwidth]{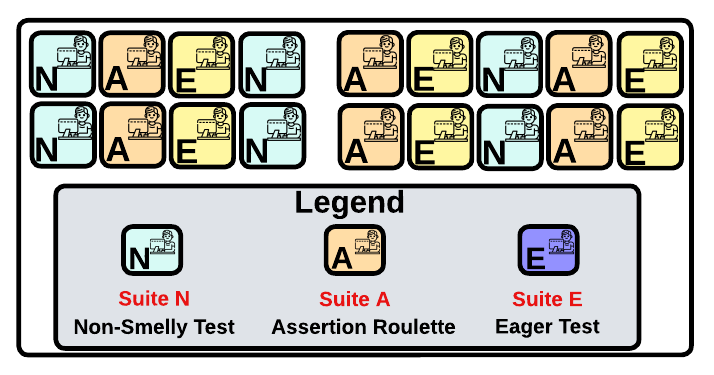}
   \vspace{-0.5cm}
   \caption{Students arrangement in the classroom.}
 \label{fig:Student_Distrubtions}
 \end{figure}


\subsection{Participants}

The controlled experiment was carried out in two semesters and involved 96 undergraduate students. 
These participants were enrolled in a software engineering class. Students were asked to complete the two sessions to obtain extra credit, but they were given the option to choose to have their data examined as part of this study. Based on that consent, out of 56 participants from semester one and 65 from semester two, we collected data from 45 participants in the first semester and 51 participants in the second semester. 
Figure \ref{fig:Student_Distrubtions} depicts each group's final arrangement in the classroom, and Table \ref{tab:participants} contains the distribution of participants in each category.

\begin{table*}
\centering
\caption{Summary of the number of participants in the study.}
\label{tab:participants}
\vspace{-0.2cm}

\begin{adjustbox}{width=0.8\textwidth,center}
\begin{tabular}{|c|c|c|c|c|c|c|}
\hline
\rowcolor{gray!60}
  \multicolumn{1}{|c|}{\cellcolor{gray!60}} &
  \multicolumn{1}{c|}{\cellcolor{gray!60}} &
  \multicolumn{1}{c|}{\cellcolor{gray!60}} &
  \multicolumn{1}{c|}{\cellcolor{gray!60}} &

  \multicolumn{3}{c|}{\cellcolor{gray!60}\textbf{Categories}} \\
\rowcolor{gray!60}

  \multicolumn{1}{|c|}{\multirow{-2}{*}{\cellcolor{gray!60}\textbf{\# of Semesters}}} &
  \multicolumn{1}{c|}{\multirow{-2}{*}{\cellcolor{gray!60}\textbf{\# of Students}}} &
  \multicolumn{1}{c|}{\multirow{-2}{*}{\cellcolor{gray!60}\textbf{\# of Eliminated Students}}} &
  \multicolumn{1}{c|}{\multirow{-2}{*}{\cellcolor{gray!60}\textbf{\# of Participated Students}}} &

  \multicolumn{1}{c|}{\cellcolor{gray!60}\textbf{Non-Test Smell}} &
  \multicolumn{1}{c|}{\cellcolor{gray!60}\textbf{Assertion Roulette}} &
  \multicolumn{1}{c|}{\cellcolor{gray!60}\textbf{Eager Test}}\\ \hline

\textbf{Semester One} &
  56 &
  11 &
  45 &
  14 &
  16 &
  15  \\ \hline
\rowcolor{gray!30}
\textbf{Semester Two} &
  65 &
  14 &
  51 &
  15 &
  17 &
  19  \\ \hline
\rowcolor{gray!30}
\cellcolor{blue!30}\textbf{Total} &
\cellcolor{blue!15}  \textbf{121} &
\cellcolor{blue!15}  \textbf{25} &
\cellcolor{blue!15}  \textbf{96} &
\cellcolor{blue!15}  \textbf{29} &
\cellcolor{blue!15}  \textbf{33} &
\cellcolor{blue!15}  \textbf{34}  \\ \hline

\end{tabular}%
\vspace{-0.3cm}
\end{adjustbox}
\vspace{-0.3cm}
\end{table*}

Upon the experiment's completion, there were 29 participants in Group \textit{N} (\textit{Non-Smelly Test}), 33 participants in Group \textit{A} (\textit{Assertion Roulette}), and 34 participants in Group \textit{E} (\textit{Eager Test}). 


\begin{lstlisting}[caption=Example of test case source code for Eager Test suite.,label=Listing:Group_C, firstnumber = last, escapeinside={(*@}{@*)},escapechar=!]
public class Calculator_ESTest extends Calculator_ESTest_scaffolding {
    @Test(timeout = 4000)
    public void test00()  throws Throwable {
        Calculator calculator0 = new Calculator();
        double[] doubleArray0 = new double[2];
        double double0 = calculator0.sum(doubleArray0);
        assertEquals(0.0, double0, 0.01);

        double[] doubleArray1 = new double[2];
        doubleArray1[0] = (-1.0);
        double double1 = calculator0.sum(doubleArray1);
        assertEquals((-1.0), double1, 0.01);

        double[] doubleArray2 = new double[2];
        double double2 = calculator0.subtract(doubleArray2);
        assertEquals(0.0, double2, 0.01);

        double[] doubleArray3 = new double[5];
        doubleArray3[2] = 93;
        double double3 = calculator0.subtract(doubleArray3);
        assertEquals((-93), double3, 0.01);
    }
}  
\end{lstlisting}


\subsection{Data Collection}

Data collection, in this study, is two-fold. First, we conducted a controlled experiment to determine how much time each participant incurred debugging the given source code. The participants began the lab session at the same time and submitted their corresponding source code after completing the debugging task. Second, we created a survey to gather details on participants' experiences in relation to debugging test cases. The survey questions were made available once they had finished identifying and fixing the bugs in the source code. Google Forms\footnote{https://www.google.com/forms/about/} was used to supply the participants with survey questions and collect the data. Two multiple-choice questions and one open-ended question were included in the questionnaire. 

\subsection{Survey}

The initial survey had eleven questions. Then, it was revised to eliminate questions that were found repetitive, irrelevant, or confusing. This revision reduced the number of questions to nine. The pilot study of the four undergraduate students revealed concerns about the length of the survey, the redundancy of some questions, and the need for logical arrangement. We reduced the nine questions to three accordingly. The final survey contains three questions—two multiple-choice and one open-ended—that can be seen in Table \ref{tab:SurveyQuestions}. 

Survey questions were made for extra credit and only for the students who participated in debugging the code. The survey respected data privacy and protection guidelines. For instance, we protect the privacy of the respondents who participated in the study by anonymizing all responses. Additionally, passwords were used to secure the researcher's laptop with all the research materials, including participant responses. Furthermore, respondents' consent was obtained prior to participation for the utilization of their data for research purposes. 

\begin{table}
\centering
\caption{Set of survey questions.}
\label{tab:SurveyQuestions}
\vspace{-0.2cm}
\begin{adjustbox}{width=0.5\textwidth,center}
\begin{tabular}{|l|c|}\hline

\rowcolor{gray!60}
  \multicolumn{1}{|c}{\cellcolor{gray!60}\textbf{Question}} &
  \multicolumn{1}{|c|}{\cellcolor{gray!60}\textbf{Type}}   \\ \hline
  

  \multicolumn{1}{|l}{Were you able to fix all the errors detected by the test cases?} &
  \multicolumn{1}{|c|}{Multiple Choice}   \\ \hline
  \multicolumn{1}{|l}{\cellcolor{gray!30}The process of finding the errors detected by the test cases } &
  \multicolumn{1}{|c|}{\cellcolor{gray!30}}   \\
  \multicolumn{1}{|l}{\cellcolor{gray!30}was smooth and easy.} &
  \multicolumn{1}{|c|}{\cellcolor{gray!30}Multiple Choice}   \\ \hline
  \multicolumn{1}{|l}{If you agreed or strongly agreed with the previous question, } &
  \multicolumn{1}{|c|}{}   \\
    \multicolumn{1}{|l}{please explain why} &
  \multicolumn{1}{|c|}{Open-ended}   \\
  \hline

\end{tabular}%
\end{adjustbox}
\vspace{-0.3cm}
\end{table}


\section{Study Results}
\label{sec:ResultsAndAnalysis}

In this section, we present the impact of \textit{Assertion Roulette} and \textit{Eager Test} on students' debugging skills (Section \ref{RQ}) and their experience with the debugging process for each test suite (Section \ref{survey}). Further, we discuss the results of suite test for each group \textit{N}, \textit{E}, and \textit{A}. 


\subsection{Experiment Results}
\label{RQ}



\vspace{-0.4cm}
\begin{box2}
\textbf{RQ: To what extent do \textit{Assertion Roulette} and \textit{Eager Test} impact the time spent by participants in debugging failing test cases?}
\end{box2}
\vspace{-0.4cm}

\textbf{Method.} The answer to this RQ, Figure \ref{fig:Boxplot} reports \textit{debugging time} boxplots of each group. To test the significance of the difference between each pair of group values, we use the Mann-Whitney U test, a non-parametric test that checks continuous or ordinal data for a significant difference between two independent groups. Our hypothesis is formulated to test whether group \textit{A} values are significantly higher than group \textit{N}. The difference is considered statistically significant if the p-value is less than 0.05. The same test is repeated for (group \textit{E}, group \textit{N}) and (group \textit{A}, group \textit{E}) pairs.

\textbf{Results.} As shown in Figure \ref{fig:Boxplot}, group \textit{A} values are significantly higher than the values of group \textit{N} (\textit{i.e.}, p$<$ 0.05). Similarly, group \textit{E} values are significantly higher than the values of group \textit{N} (\textit{i.e.}, p$<$ 0.05). The two pairwise comparisons indicate how students who were using either Suite \textit{A} or Suite \textit{E} have spent a significantly larger amount of time locating the errors, in comparison with students who were using Suite \textit{N}.


\textbf{Observations.} The time spent to locate an error differs depending on the test suite reporting it. For instance, when using Suite \textit{N}, each failing test method contains only one failing assert statement. This failing statement would eventually indicate to the student the inconsistency between the expected value and the actual value of the method under test. The students would then investigate the corresponding method. When the error is found and then fixed, not only the investigated testing method would pass, but also any other test methods that were failing for the same reason.
On the other hand, when using Suite \textit{A}, each failing test method contains multiple assert statements, in which a subset is failing with more than one assert statement to be investigated. The students tend to examine them to understand the common cause for their failure before switching to the method under test to search for the error. Therefore, the investigation of the multiple asserts seems to increase the debugging time, and therefore, the \textit{Assertion Roulette} is negatively impacting the students' debugging process. Our observation brings an alternate perspective with respect to the recent findings of Bai et al. \cite{bai2022assertion}, who conjectures that \textit{Assertion Roulette} may no longer be considered a code smell. In fact, Bai et al. \cite{bai2022assertion} demonstrated, through a controlled classroom experiment, that \textit{Assertion Roulette} does impact neither the frequency of testing nor the accuracy of the test cases. We argue that \textit{Assertion Roulette} is a program comprehension problem. Therefore, a controlled experiment where students' programming performance and testing behaviors are measured may not reveal the negative symptoms of \textit{Assertion Roulette}.           
As for the \textit{Eager Test}, the smell is caused by a test case involving multiple production methods, thereby increasing coupling between the test and production code. Such situations result in participants reviewing multiple production methods and continually switching between multiple code files as part of their troubleshooting task, which potentially increases their cognitive load and debugging time.

\vspace{-0.3cm}
\begin{box1}
   \textbf{Summary:} The presence of test smells, namely \textit{Assertion Roulette} and \textit{Eager Test}, in the test suite cause participants to spend more time troubleshooting test case failures, in comparison with performing the same debugging process using a non-smelly suite.
\end{box1}
\vspace{-0.3cm}

\pgfplotsset{
    box plot/.style={
        /pgfplots/.cd,
        blue,
        fill=blue!20,
        only marks,
        mark=-,
        mark size=1em,
        /pgfplots/error bars/.cd,
        y dir=plus,
        y explicit,
    },
    box plot box/.style={
        /pgfplots/error bars/draw error bar/.code 2 args={%
            \draw  ##1 -- ++(1em,0pt) |- ##2 -- ++(-1em,0pt) |- ##1 -- cycle;
        },
        /pgfplots/table/.cd,
        y index=2,
        y error expr={\thisrowno{3}-\thisrowno{2}},
        /pgfplots/box plot
    },
    box plot top whisker/.style={
        /pgfplots/error bars/draw error bar/.code 2 args={%
            \pgfkeysgetvalue{/pgfplots/error bars/error mark}%
            {\pgfplotserrorbarsmark}%
            \pgfkeysgetvalue{/pgfplots/error bars/error mark options}%
            {\pgfplotserrorbarsmarkopts}%
            \path ##1 -- ##2;
        },
        /pgfplots/table/.cd,
        y index=4,
        y error expr={\thisrowno{2}-\thisrowno{4}},
        /pgfplots/box plot
    },
    box plot bottom whisker/.style={
        /pgfplots/error bars/draw error bar/.code 2 args={%
            \pgfkeysgetvalue{/pgfplots/error bars/error mark}%
            {\pgfplotserrorbarsmark}%
            \pgfkeysgetvalue{/pgfplots/error bars/error mark options}%
            {\pgfplotserrorbarsmarkopts}%
            \path ##1 -- ##2;
        },
        /pgfplots/table/.cd,
        y index=5,
        y error expr={\thisrowno{3}-\thisrowno{5}},
        /pgfplots/box plot
    },
    box plot median/.style={
        /pgfplots/box plot
    }
}

\begin{figure}[h!]
     \centering
     
\begin{adjustbox}{width=0.4\textwidth}

\begin{tikzpicture}
\begin{axis} [enlarge x limits=0.2,xtick=data,xticklabels={Non-Smelly Test,Assertion Roulette,Eager Test},ylabel=Time (Minutes), xlabel=Test Suites,ymin=0,ymax=120
]
    \addplot [box plot median] table {testdata.dat};
    \addplot [box plot box] table {testdata.dat};
    \addplot [box plot top whisker] table {testdata.dat};
    \addplot [box plot bottom whisker] table {testdata.dat};
\end{axis}
\end{tikzpicture}

\end{adjustbox}
\vspace{-0.3cm}
 \caption{Distribution of the time spent by participants to detect and debug each test suite.}
\label{fig:Boxplot}
\end{figure}
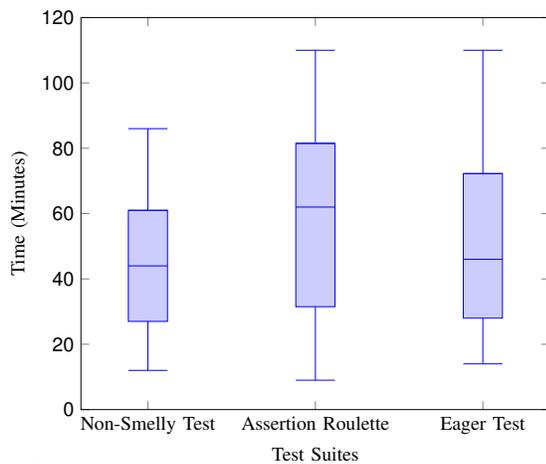

\begin{figure*}[h]
\centering
\caption*{The process of finding the errors detected by the test cases was smooth and easy?}
\resizebox{\textwidth}{!}{
\begin{tabular}{ccc}

\begin{bchart}[step=20,max=100,width=5cm, unit=\%]
\bcbar[label=Strongly Agree]{55}
\bcbar[label=Agree]{28}
\bcbar[label=Neutral]{14}
\bcbar[label=Disagree]{0}
\bcbar[label=Strongly Disagree]{3}

\end{bchart}
&
\begin{bchart}[step=20,max=100,width=5cm, unit=\%]
\bcbar[label=Strongly Agree]{40}
\bcbar[label=Agree]{27}
\bcbar[label=Neutral]{18}
\bcbar[label=Disagree]{9}
\bcbar[label=Strongly Disagree]{6}

\end{bchart}
&
\begin{bchart}[step=20,max=100,width=5cm, unit=\%]
\bcbar[label=Strongly Agree]{50}
\bcbar[label=Agree]{32}
\bcbar[label=Neutral]{15}
\bcbar[label=Disagree]{3}
\bcbar[label=Strongly Disagree]{0}

\end{bchart}
\\
\textbf{(A) Non-Smelly Test} & \textbf{(B) Assertion Roulette}   & \textbf{(C) Eager Test}\\

\end{tabular}}
\vspace{-0.2cm}
 \caption{Distribution of students' responses to survey question \#2.}
\label{fig:Result_Q1}
\vspace{-0.4cm}
\end{figure*}


\subsection{Survey Results}
\label{survey}


\textbf{Method.} 
We also wanted to know if the debugging task was challenging or straightforward for the participants to complete. Thus, we analyzed the participants' survey responses after they submitted them. Figure~\ref{fig:Result_Q1} presents the responses by the participants divided based on each group. We asked the participants three questions.

\textbf{Results.} 
We questioned the participants, \textbf{\textit{"Were you able to fix all the errors detected by the test cases?"}}, to investigate the impact of test smells on the performance of participants' detection and debugging skills of the source code. As the proportion of successful participants finishing the task within the given time frame significantly impacts their performance, this question is essential to the generalizability of the analysis. We compiled the responses, and we verified them with the files that were submitted. We found that a total of 13 participants were unable to locate and fix the source code, with the number of participants assigned to the \textit{Assertion Roulette} test suite failing by the biggest margin ($n=7$). Three participants who were provided the \textit{Non-Smelly Test} test suite did not manage to complete the assignment. Similarly, 3 participants in the test suite for the \textit{Eager Test} failed to complete the task. These statistics illustrate the considerable detrimental impact of \textit{Assertion Roulette} on participants' ability to debug source code. It should be highlighted that participants were uninformed of the test smells existence in the suites.

\vspace{-0.3cm}
\begin{box1}
\textbf{Summary:} 
Although the participants were unaware of the test smells they were assigned, most of them managed to locate and fix all the issues in the source code. Additionally, participants assigned with the \textit{Assertion Roulette} test suite had the highest percentage that was unable to complete the assignment within the given time frame revealing the negative impact of this particular smell.
\end{box1}
\vspace{-0.1cm}

Figure \ref{fig:Result_Q1} showcases the response of participants regarding the second multiple choice question of the survey; \textbf{\textit{"The process of finding the errors detected by the test cases was smooth and easy?"}} Five options—Strongly Agree, Agree, Neutral, Disagree, or Strongly Disagree—were given to the participants. The majority of participants, in general, strongly agreed that the process of identifying and debugging test cases is simple and straightforward. 

Figure \ref{fig:Result_Q1} (A) corresponds to the responses received by the participants from \textit{Non-Smelly Test} test suite. It is clear that the highest ratio (55\% $n=16$) of participants agree with the smoothness and straightforwardness of the underlying debugging task. Following this, 28\% ($n=8$) of the participants selected the "\textit{Agree}" option, which makes an overall 83\% ($n=24$) of the participants who agreed with the simplicity of locating and fixing bugs in \textit{Non-Smelly Test} test suite. On the contrary, only a small portion of participants, i.e., 3\% ($n=1$), strongly disagreed hence, referring to the procedure as difficult, while none of the participants disagreed with the statement. On the other hand, 14\% ($n=4$) of the participants remained neutral to the asked question. Therefore,  it is fair to say that the absence of test smells in the test cases makes the participants' task of debugging easy to understand and simple to fix.

Figure \ref{fig:Result_Q1} (B) showcases the responses received by the participants of \textit{Assertion Roulette} test suite to the survey question \#2. Unlike the previously discussed case, only 67\% ($n=22$) participants of \textit{Assertion Roulette} test suite coincided with the process of locating and fixing the bugs to be easy and smooth, among which 40\% ($n=13$) participants were in strong agreement with the statement and 27\% ($n=9$) of the participants agreed. Moreover, the ratio of participants who disagreed and strongly disagreed with the stated question amounted to 15\% ($n=5$), among which 9\% ($n=3$) disagreed and 6\% ($n=2$) of the participants strongly disagreed. These numbers may have formed due to the test cases' inclusion of numerous untitled assertions, which confused the participants. The participants spent more time locating the assertion that caused the test case to fail, making it more difficult to handle. Therefore, the most difficult test smell to address in the test cases can be ruled out as \textit{Assertion Roulette}. 

Figure \ref{fig:Result_Q1} (C) presents the statistical data regarding participants' responses to survey question \#2 in terms of \textit{Eager Test} test suite. Overall, 82\% ($n=28$) of the underlying participants voted in conjunction with the stated question. Where, 50\% ($n=17$) strongly agreed with the ease and simplicity of the process and 32\% ($n=11$) of the participants responded with "\textit{Agree}" option. On the other hand, only 3\% ($n=1$) of the participants disagreed with the process of locating and fixing bugs in \textit{Eager Test} test suit being easy and smooth. In \textit{Eager Test} test suit, the test cases from \textit{Non-Smelly Test} and \textit{Assertion Roulette} test suites were merged into 9 test cases. Moreover, this smell induces multiple production codes when one method is invoked, causing a significant impact on the debugging skills of the participants. However, participants found it comparatively easy to handle as they were able to follow up on the logic of the operation and locate the bug in the source code.

\textbf{Observations.} 
Overall, students from all groups exhibit similar levels of satisfaction with their testing experience, with a slight increase for those using non-smelly tests, and those using \textit{Eager Tests}. None of the participants have reported any issues they experienced, despite the smelliness of Suite \textit{A} and \textit{E}. We argue that \textit{Assertion Roulette} and \textit{Eager Test} are hard to be sought as problematic, as they are intuitive by nature. This is aligned with their high frequency in open source systems, according to recent studies \cite{bavota2012empirical,grano2019scented,peruma2019CASCON,Spadini2020MSR}.




\vspace{-0.1cm}
\begin{box1}
   \textbf{Summary:} \textit{Non-Smelly Test} and \textit{Eager Test} test suites were equally rated to be simple and straightforward to handle. In contrast, \textit{Assertion Roulette} test smell was rendered to make the process difficult for the participants. 
\end{box1}

Finally, we supplied the participants with an open-ended question: \textbf{\textit{If you agreed or strongly agreed with the previous question, please explain why.}} 
Responses received from group \textit{N} participants show their high satisfaction with the overall testing experience: 

\begin{quote}
        \faCommenting \hspace{0.08cm}\textit{\textbf{Comment 1: }``The structure was easier to diagnose with each test cases. Once I was able to diagnose where the errors were, it got easier to clear all the errors.''} 
\end{quote}

\begin{quote}
        \faCommenting \hspace{0.08cm}\textit{\textbf{Comment 2: }``Once I found which test case was which, it was easy to find what function they were using (after I realized the equate function was a built in one). That in turn made it easy to figure out that the error was limited in scope to that function alone [...]''} 
\end{quote}

Moreover, some participants who were not previously familiar with the Java programming language expressed a positive experience, as one of the participants said:

\begin{quote}
        \faCommenting \hspace{0.08cm}\textit{\textbf{Comment 3: }``Haven't touched java before, once I understood what was going on, wasn't difficult to find''} 
\end{quote}


Responses received from group \textit{A} participants did not differ from the previous group, as students expressed their satisfaction with the debugging process:

\begin{quote}
        \faCommenting \hspace{0.08cm}\textit{\textbf{Comment 4: }`` [...] Because I can see the expected result and the actual result and trace back where the code is wrong.''} 
\end{quote}






Responses received from group \textit{E} participants did not deviate from the previous ones, as students positively described their debugging:


\begin{quote}
        \faCommenting \hspace{0.08cm}\textit{\textbf{Comment 5: }``The print out statement and the expected output helped me identified where in the code to look and what I needed to fix.''} 
\end{quote}

\begin{quote}
        \faCommenting \hspace{0.08cm}\textit{\textbf{Comment 6: }``The test cases pointed out some functions being used and gave the expected and actual values. This helped me to pinpoint what was wrong and where.''} 
\end{quote}

\begin{quote}
        \faCommenting \hspace{0.08cm}\textit{\textbf{Comment 7: }``I found it easy to  to find the issues because the code was well formatted and commented. When I saw the error was with subtraction I went to the subtraction function. There I read the comments on what each part of the code was supposed to do, when it was different I just changed the code to do what the comments said.''} 
\end{quote}

Among these positive comments, We note how \textit{comment 6} has mentioned the test of more than one method without necessarily considering it to be problematic. It is apparent that the successful fix of all errors raised a sense of accomplishment among students and positively influenced them.  



\vspace{-0.3cm}
\begin{box1}
   \textbf{Summary:} \textit{Assertion Roulette}, and \textit{Eager Test} smells are transparent to students as long as they are successful in locating and fixing errors. 
\end{box1}
\vspace{-0.3cm}

\section{Discussion}
\label{sec:Discussions}

The analysis of the current investigation has revealed significant information regarding the impact of \textit{Assertion Roulette} and \textit{Eager Test} on the participants' debugging time.

As we previously discussed, Bai et al. \cite{bai2022assertion} advocates for \textit{Assertion Roulette} to be no longer considered as test smell. However, the aforesaid statement is not consistent with our findings. Our experiment revealed that participants spent significantly more time sorting the assertions that are stacked in the \textit{Assertion Roulette} test methods. Furthermore, we witnessed a similar pattern of longer debugging time, when students use \textit{Eager Test} test methods to locate errors.

Therefore, educators need to raise the students' awareness of writing smell-free test cases. Students need to be taught how to avoid writing multiple asserts under the same test methods, or to test multiple production methods, using the same test method. In this context, Buffardi et al. examined students' test methods on their assignments, and indicated potential problems in their unit tests. In fact, the top three common patterns detected in their test methods: multiple member function calls, multiple assertions, and conditional logic \cite{buffardi2021unit}. Thereby, taking an action to educate students on how to avoid these anti-patterns, would decrease the early propagation of these smells.

Additionally, the current research discovered that test cases with descriptive names and commented asserts increase their readability \cite{scalabrino2016improving}. Thus, we encourage students to develop the practice of documenting their test methods.
 

\subsection{Lessons Learned:}
Throughout the experiment, several lessons were learned, and perceptual observations were conducted.

\faPaperclip \hspace{0.08cm}\textbf{Lesson \#1:} In addition to teaching students about design and code smells, academia must also instill in students the importance of writing quality test cases, specifically test smells and the harm caused by the introduction and existence of test smells in the systems code base. Furthermore, teaching students about code reviews should not be limited to the production code but should also include the test suites, as such code is vital to the system's overall quality.  

\faPaperclip \hspace{0.08cm}\textbf{Lesson \#2:} The research community has produced tools to aid developers with detecting (and, in some cases) correcting test smells for various programming languages, and testing frameworks \cite{aljedaani2021test}. These tools have been utilized in multiple empirical studies and have been effective in their detection mechanism. Academia should promote using these tools in the classroom to better equip students with means of evaluating the quality of the test cases they produce for class assignments. In addition, the automatic detection of test smells will help students to troubleshoot issues much faster.

\faPaperclip \hspace{0.08cm}\textbf{Lesson \#3:} Even though most research on test smells focuses on Java systems, test smells are not unique to JUnit-based test suites. Therefore, the research and academic community should invest in exploring the types of test smells that are unique to specific programming languages and paradigms and passing that knowledge to students in the classroom so that they are better prepared when entering the workforce.

\section{Threats to Validity}
\label{sec:ThreatsValidity}

The applicability of the findings in this research is susceptible to a number of threats.

\textbf{Internal Validity.} 
The students who participated in this experiment were from the same university, making the participant pool relatively coherent. To mitigate this, we run our experiments in two semesters. Another pertinent threat relates to the choice of the project under test. The complexity of the project features may require advanced debugging skills, which may crate a significant overhead in our experiment. To mitigate this issue, we developed a calculator application with eight operations. The choice of the calculator ensures its intuitiveness to the students. Although some students might not be familiar with the source code, we gave them detailed instructions and documentation to familiarize themselves with the project and source code in order to minimize the threat and potential misunderstandings.

\textbf{External Validity.}
This study is exclusively focused on two test smells. However, given the prevalence of test smells, there may be a chance that we missed an important smell that is typically written by developers. To mitigate this threat, we reviewed various research papers that have conducted similar types of experiences. We concluded that the most popular and frequent smells are \textit{Assertion Roulette}, and \textit{Eager Test}. 
Moreover, it is essential to replicate the study with a broader and more varied range of test smells along with a wider range of codebases. 

\section{Conclusion and Future Work}
\label{sec:Conclusion}

In this study, we explored the impact of test smells on a student's program comprehension ability. The results indicate that the participants in our study who were assigned test suites containing test smells took more time to complete the assigned task of fixing errors in the production code than those given test suites without test smells. Furthermore, the smell \textit{Assertion Roulette} had a greater impact on troubleshooting and debugging than the \textit{Eager Test} smell. For future work, we plan to conduct a similar study with additional test smell types to determine how these new smells impact a student's ability to comprehend code when performing maintenance activities.

\section*{Verifiability and Replicability}
To enable full verifiability and replicability, our experimental materials are available\footnote{https://wajdialjedaani.github.io/testsmellstd/} \cite{ProjectWebiste}.

\bibliographystyle{abbrv}
\bibliography{main}

\end{document}